\documentclass{kluwer}    

\newdisplay{guess}{Conjecture}

\begin{document}                                                                                   
\begin{article}
\begin{opening}         
\title{Properties of Elastic Waves in a non-Newtonian (Maxwell)
Fluid-Saturated Porous Medium}
\author{David \surname{Tsiklauri}\email{tsikd@astro.warwick.ac.uk}}
\institute{Physics Department, University of Warwick, Coventry, CV4 7AL, UK}
\author{Igor \surname{Beresnev}\email{beresnev@iastate.edu}}
\institute{Department of Geological 
and Atmospheric Sciences,
Iowa State University, 253 Science I, Ames, IA 50011-3212,
U.S.A.}
\runningauthor{Tsiklauri \& Beresnev}
\runningtitle{Elastic Waves in a Maxwell fluid-Saturated matrix}
\date{\today}
\begin{abstract}
The present study investigates novelties
brought about into the classic Biot's theory of 
propagation of elastic waves in a fluid-saturated
porous solid by inclusion of non-Newtonian effects
that are important, for example, for hydrocarbons.
Based on our previous results (Tsiklauri \& Beresnev: 2001, 
{\it Phys. Rev. E}, {\bf 63}, 046304),
we have investigated the propagation of 
rotational and dilatational elastic waves, through
calculating their phase velocities and attenuation
coefficients as a function of frequency.
We found that the replacement of an ordinary Newtonian fluid
by a Maxwell fluid in the fluid-saturated
porous solid results in: 
(a) an overall increase of the phase velocities of both the rotational  
and dilatational waves. With the increase of frequency these
quantities tend to a fixed, higher, 
as compared to the Newtonian limiting case, level
which is not changing with the decrease of the Deborah number
$\alpha$. 
(b) the overall decrease of the attenuation coefficients of both the 
rotational and dilatational waves. With the increase of frequency these
quantities tend to a progressively lower,
as compared to the Newtonian limiting case, levels
as $\alpha$ decreases. 
(c) Appearance of oscillations in all physical quantities 
in the deeply non-Newtonian regime.
\end{abstract}
\end{opening}

\section{Introduction}

Apart from fundamental interest,
there are at least three major reasons to study the
dynamics of fluid in porous media under oscillatory pressure
gradient and oscillating pore walls, as well as
to investigate propagation of elastic waves
in porous media.

First, in petroleum geophysics, regional exploration seismology
needs direct methods of discovering oil-filled bodies of rock,
and these should be based on models of propagation of elastic
waves in porous media with {\it realistic fluid rheologies} 
(Carcione \& Quiroga-Goode, 1996).

Second, the investigation of the
dynamics of fluid in porous media under oscillatory pressure
gradient  
is of prime importance for the recently emerged technology
of acoustic stimulation of oil reservoirs 
(Beresnev \& Johnson, 1994), (Drake \& Beresnev, 1999).
It is known that the natural pressure in an oil reservoir 
generally yields no more than approximately 10 percent oil recovery.  
The residual oil is difficult to produce due to its naturally 
low mobility, and the enhanced oil recovery operations are 
used to increase production.  It has been experimentally proven
that there is a substantial increase in the net fluid flow 
through porous space if the latter is treated with elastic waves 
(Beresnev \& Johnson, 1994), (Drake \& Beresnev, 1999).  

Third, in the environment conservation, treatment of ground water 
aquifers contaminated by organic liquids, such as
hydrocarbons, by elastic waves proved to be successful 
for quick and efficient clean up 
(Beresnev \& Johnson, 1994), (Drake \& Beresnev, 1999).

A quantitative theory of propagation 
of elastic waves in a fluid-saturated
porous solid was formulated in the classic papers by Biot (Biot, 1956a,b).
One of the major findings of Biot's work was 
that there is
a breakdown in Poisseuille flow above a certain characteristic frequency
specific to the fluid-saturated porous material. 
Biot theoretically studied this phenomenon by
considering the flow of a viscous fluid in a tube with
longitudinally oscillating walls under an oscillatory pressure
gradient. 
In Biot's theory, 
the two-phase material is considered as a continuum and
the microscopic, pore-level effects are ignored.
As a reminder, 
the theory assumes that: (a) the wavelength is large 
with respect to
the dimensions of pores in order to make continuum
mechanics applicable; this also implies that scaterring
dissipation is negligible; (b) the displacements are small, therefore
the macroscopic strain tensor is related to them by the
lowest second-order approximation; (c) the liquid phase is
continuous, such that the pores are connected and 
isolated pores are treated as part of solid matrix; and (d)
the permeability is isotropic and the medium is fully
saturated.

Biot demonstrated the existence of the two kinds of compressional
waves in  a fluid-saturated porous medium: the fast wave
for which the solid and fluid displacements are in phase,
and the slow wave for which  the displacements are out
of phase. At low frequencies, the medium does not
support the slow wave as it becomes diffusive. 
On the other hand, at
high frequencies, tangential slip takes place, inertial effects
dominate, and the slow wave becomes activated.

Biot's theory can be used to describe interaction of fluid-saturated 
solid with the sound for a classic Newtonian fluid;
however, oil and other hydrocarbons often exhibit significant
non-Newtonian behavior.
In paper I  (Tsiklauri \& Beresnev, 2001), 
we have incorporated non-Newtonian effects
into the classical theory of Biot (Biot, 1956a,b).
Using the recent results of del Rio {\it et al.} (1998),
who presented a
study of enhancement in the dynamic response of a viscoelastic 
(Maxwell) fluid
flowing through a stationary (non-oscillating) tube under the effect of
an oscillatory pressure gradient,
we have combined their theory with the effect of the acoustic
oscillations of the walls of the tube introduced by Biot (Biot, 1956a,b), thus
providing a complete description of the interaction
of Maxwell fluid, filling the pores, with acoustic waves.
We have generalized the expression for the function $F(\kappa)$, 
which measures the deviation from
Poisseuille flow friction as a function of frequency 
parameter $\kappa$ (Tsiklauri \& Beresnev, 2001).
As a next step, in the present work we investigate the propagation of 
rotational and dilatational elastic waves
through the porous medium filled with Maxwell fluid, by
calculating their phase velocities and attenuation
coefficients as a function of frequency.

This paper is organized as follows: we formulate
theoretical basis for our numerical calculations in section 2.
In sections 3 and 4 we study numerically properties of the 
rotational and dilatational  elastic waves, respectively, 
and, finally, in Section 5 
we close with a discussion of our main results.

\section{Theory}

The theory of propagation 
of elastic waves in a fluid-saturated
porous solid was formulated  by Biot (Biot, 1956a,b).
He demonstrated that the general equations  
which govern propagation
of rotational and dilatational high-frequency waves in a 
fluid-saturated porous medium are the same as in the low-frequency 
range provided the viscosity is replaced by its
effective value as a function of frequency.
In practice, it means replacing the resistance
coefficient $b$ by $bF(\kappa)$.

The equations describing dynamics of the rotational waves are (Biot, 1956a,b)
$$
{{\partial^2}\over{\partial t^2}}(\rho_{11}\vec \omega + \rho_{12}\vec
\Omega)
+b F(\kappa){{\partial}\over{\partial t}}(\vec \omega - \vec \Omega)=
N \nabla^2 \vec \omega,
\eqno(1)
$$
$$
{{\partial^2}\over{\partial t^2}}(\rho_{12}\vec \omega + \rho_{22}\vec
\Omega)
- b F(\kappa){{\partial}\over{\partial t}}(\vec \omega - \vec \Omega)=0,
\eqno(2)
$$
where, $\rho_{11},\rho_{12}$ and $\rho_{22}$ are mass density
parameters for the solid and fluid and their inertia coupling; 
$\vec \omega= \mathrm{curl}\; \vec u$ and $\vec \Omega =\mathrm{curl}\; 
\vec U$  describe 
rotations of solid and fluid with $\vec u$ and $\vec U$ 
being their displacement 
vectors, while
the rigidity of the solid is represented by the modulus $N$.
Substitution of a plane rotational
wave of the form
$$
\omega=C_1 e^{i(l x + \chi t)}, \;\;\;
\Omega=C_2 e^{i(l x + \chi t)},
\eqno(3)
$$
into Eqs.(1) and (2) allows us to obtain a characteristic equation
$$
{{Nl^2}\over{\rho \chi^2}}=E_r -i E_i,
\eqno(4)
$$
where $l$ is wavenumber, $\chi=2 \pi f$ is  wave cyclic frequency,
 $\rho=\rho_{11}+2 \rho_{12}+\rho_{22}$ is the mass
density of the bulk material.

The real and imaginary parts of  Eq.(4) can be written
as
$$
E_r={{(\gamma_{11}\gamma_{22}-\gamma_{12}^2)(\gamma_{22}
+\epsilon_2)+\gamma_{22}\epsilon_2 
+\epsilon_1^2+\epsilon_2^2}\over{(\gamma_{22}+\epsilon_2)^2+\epsilon_1^2}},
\eqno(5)
$$
and
$$
E_i={{\epsilon_1 (\gamma_{12}+\gamma_{22})^2}
\over{(\gamma_{22}+\epsilon_2)^2+\epsilon_1^2}},
\eqno(6)
$$
where $\gamma_{ij}=\rho_{ij}/ \rho$,
$\epsilon_1=(\gamma_{12}+\gamma_{22})
(f_c/f)\,Re[F(\kappa)]=$ \\
$(\gamma_{12}+\gamma_{22})
(f_c/f)\,Re[F(\delta \sqrt{f/f_c})]$,
$\epsilon_2=(\gamma_{12}+\gamma_{22})
(f_c/f)\,Im[F(\kappa)]=(\gamma_{12}+\gamma_{22})
(f_c/f)\,Im[F(\delta \sqrt{f/f_c})]$.
The function $F(\kappa)$ was written here more conveniently 
as a function of frequency $f$ , i.e.
$F(\kappa)=F(\delta \sqrt{f/f_c})$ (Biot, 1956a,b),
where $\delta$ is a factor dependent on pore
geometry. For the hollow cylinder-like pores, 
$\delta=\sqrt{8}$ (Biot, 1956a,b) and we use this value
throughout the paper. $f_c$ is the critical
frequency above which the Poisseuille flow breaks down,
and it equals  $b/(2 \pi \rho_2)=b/(2 \pi \rho(\gamma_{12}+
\gamma_{22}))$. 

In order to obtain phase velocity and attenuation coefficient
of the rotational waves, we put $l=Re[l]+i Im[l]$. Thus, the phase
velocity is then $v_r= \chi / |Re[l]|$. Introducing
a reference velocity as $V_r=\sqrt{N/ \rho}$, we obtain the
dimensionless phase velocity as
$$
{{v_r}\over{V_r}}={{\sqrt{2}}\over{\left[ \sqrt{E_i^2+E_r^2}
+E_r\right]^{1/2}}}.
\eqno(7)
$$

To obtain the attenuation coefficient
of the rotational waves, we introduce
a reference length, $L_r$,  defined as
$L_r=V_r/(2 \pi f_c)$. The length $x_a$
represents the distance over which the
rotational wave amplitude is attenuated by a factor
of $1/e$. Therefore we can construct the dimensionless
attenuation coefficient as $L_r/x_a$,
$$
{{L_r}\over{x_a}}={{f}\over{f_c}}
{{\left[\sqrt{E_i^2+E_r^2} -E_r\right]^{1/2}}\over{\sqrt{2}}}.
\eqno(8)
$$

The equations describing dynamics of the dilatational waves are (Biot, 1956a,b)
$$
\nabla^2(Pe+Q \epsilon)=
{{\partial^2}\over{\partial t^2}}(\rho_{11} e + \rho_{12} \epsilon)
+b F(\kappa){{\partial}\over{\partial t}}(e - \epsilon),
\eqno(9)
$$
$$
\nabla^2(Qe+R \epsilon)=
{{\partial^2}\over{\partial t^2}}(\rho_{12} e + \rho_{22} \epsilon)
-b F(\kappa){{\partial}\over{\partial t}}(e - \epsilon),
\eqno(10)
$$
where, $P,Q$ and $R$ are the elastic coefficients, $e=\mathrm{div} \;
\vec u$ and 
$\epsilon=\mathrm{div} \; \vec U$ are the divergence of solid 
and fluid displacements.
Again, substitution of a plane dilatational wave of the form
$$
e=C_1 e^{i(l x + \chi t)}, \;\;\;
\epsilon=C_2 e^{i(l x + \chi t)},
\eqno(11)
$$
into Eqs.(9) and (10) allows us to obtain a characteristic equation
$$
(z-z_1)(z-z_2)+ i M(z-1)=0,
\eqno(12)
$$
where $z=l^2 V_c^2/ \chi^2$, $V_c^2=(P+R+2Q)/ \rho$ represents
the velocity of a dilatational wave when the
relative motion between fluid and solid is absent,
$z_{1,2}=V_c^2/V_{1,2}^2$ with $V_{1,2}$ being the velocities
of the purely elastic waves with subscripts 1,2 referring to
the two roots of Eq.(12), and finally 
$M=(\epsilon_1 + i \epsilon_2)/(\sigma_{11} \sigma_{22}- \sigma_{12}^2)$
with $\sigma_{11}=P/(P+R+2Q)$, $\sigma_{22}=R/(P+R+2Q)$ and 
$\sigma_{12}=Q/(P+R+2Q)$.

Eq.(12) has two complex roots $z_I$ and $z_{II}$.
Phase velocities of the two kinds of dilatational waves
can be defined as 
$$
{{v_I}\over{V_c}}={{1}\over{Re[\sqrt{z_I}]}},
\;\;\; 
{{v_{II}}\over{V_c}}={{1}\over{Re[\sqrt{z_{II}}]}},
\eqno(13)
$$
while the corresponding attenuation coefficients can be
also introduced as
$$
{{L_c}\over{x_I}}={Im[\sqrt{z_I}]}{{f}\over{f_c}},
\;\;\; 
{{L_c}\over{x_{II}}}={Im[\sqrt{z_{II}}]}{{f}\over{f_c}}.
\eqno(14)
$$

In paper I, we generalized Biot's expression
for  $F(\kappa)$ to the case of a non-Newtonian
(Maxwell) fluid, which reads 
$$
F(\kappa)=-{{1}\over{4}}
{ { \kappa \sqrt{ i + \kappa^2 / \alpha}
\left[J_1(\kappa \sqrt{ i + \kappa^2 / \alpha})
/ J_0(\kappa \sqrt{ i + \kappa^2 / \alpha})\right]}
\over{(1 -i \kappa^2 / \alpha)}}
$$
$$
\left[1 - {{2 J_1(\kappa \sqrt{ i + \kappa^2 / \alpha})}\over
{\kappa \sqrt{ i + \kappa^2 / \alpha}
J_0(\kappa \sqrt{ i + \kappa^2 / \alpha})}}
\right]^{-1}.
\eqno(15)
$$
Here, $\kappa= a \sqrt{\omega / \nu}$ is the frequency
parameter, $a$ is the radius of the pore,  $\nu= \eta / \rho$
is the ratio of the viscosity coefficient to the fluid mass density, 
$J_0$ and $J_1$ are the Bessel functions, and, finally,
$\alpha$ denotes the Deborah number del Rio {\it et al.} (1998), 
which is defined as
the ratio of the characteristic time of viscous effects $t_v=a^2/ \nu$
to the 
relaxation time $t_m$, i.e.,
$\alpha= t_v / t_m =a^2/(\nu t_m)$.

Eq.(15) was derived by solving the equations of incompressible
hydrodynamics,
namely, the continuity equation, linearized momentum equation, and
rheological
equation of a Maxwell fluid, in the frequency domain for a cylindrical
tube whose walls oscillate harmonically in time.
By calculating the ratio of the total friction force exerted on the tube
wall
to the average velocity
of a Maxwell fluid, and noting that $F(\kappa)$ is proportional to this
ratio,
we generalized the classical result obtained by Biot (see details in
Paper I).

As noted in  del Rio {\it et al.} (1998), the value of the parameter $\alpha$
determines in which regime the system resides. Beyond 
a certain critical value ($\alpha_c=11.64$), the system is
dissipative, and viscous effects dominate. On the other hand, for
small $\alpha$ ($\alpha < \alpha_c$), the system
exhibits viscoelastic behavior which we call the 
non-Newtonian regime.
Note, that the Newtonian flow regime can be easily recovered from 
Eq.(15) by putting $\alpha \to \infty$.

In order to investigate the novelties brought about
into classical Biot's theory of propagation
of elastic waves in porous medium (Biot, 1956a,b) by the
inclusion of non-Newtonian effects, we have
studied the full parameter space of the problem.
We have calculated the normalized phase velocities and
attenuation coefficients for both
rotational and dilatational waves using 
our general expression for $F(\kappa)$ given by Eq. (15).

\section{Numerical Results for Propagation of Rotational Waves}

In all our numerical calculations, we have used polynomial
expansions of $J_0$ and $J_1$ with absolute error not
exceeding $10^{-6}$ \%. Thus, our calculation results are 
accurate to this order. Also, 
in order to catch a true oscillatory structure of our
solutions (see below), number of data points
in all our plots is 10000, as opposed to paper I
where only 100 points per curve were taken. 

In all forthcoming results, we calculate phase velocities and
attenuation coefficients for the case 1 from Table I
taken from  (Biot, 1956b), which is $\sigma_{11}=0.610$, 
$\sigma_{22}=0.305$, $\sigma_{12}=0.043$, $\gamma_{11}=0.500$,
$\gamma_{22}=0.500$, $\gamma_{12}=0$, $z_{1}=0.812$, and 
$z_{2}=1.674$.

We calculated normalized phase velocity
of the plane rotational waves, $v_r/V_r$, 
and the attenuation coefficient $L_r/x_{a}$ using 
our more general expression for $F(\kappa)$ 
(Maxwell fluid filling the pores)
given by Eq. (15).

In Fig. 1 we plot phase velocity 
$v_r/V_r$ as a function of frequency for the three
cases: the thick  curve corresponds to $\alpha \to \infty$
(Newtonian limit),  the dashed  curve
corresponds to a slightly sub-critical value of $\alpha=10$ 
(recall that $\alpha_c=11.64$) and thin solid curve
corresponds to the case of $\alpha=1$.
Note that the $\alpha \to \infty$ case perfectly reproduces
the curve 1 in Fig. 5 from  (Biot, 1956b). For $\alpha=10$
we notice a deviation from the classic Newtonian
behavior in the form of overall increase of phase
velocity and appearance of small oscillations
on the curve, which means that we have entered the 
non-Newtonian regime.
Note that when $\alpha=1$ the phase velocity settles at
somewhat higher value and this onset happens
already for smaller frequencies than in the
case of Newtonian fluid. Also, much more
pronounced oscillatory structure of the solution
can be observed.

Fig. 2 shows the attenuation coefficient  
$L_r/x_a$ of the rotational wave
 as a function of frequency for the three
values of $\alpha$: the thick  curve corresponds to $\alpha \to \infty$
(Newtonian limit), the dashed  curve
corresponds to a slightly sub-critical value of $\alpha=10$ 
and thin solid curve
corresponds to the case of $\alpha=1$.
Note that $\alpha \to \infty$ case  coincides with
curve 1 in Fig. 6 from  (Biot, 1956b). For $\alpha=10$,
there is a noticeable deviation from the classic Newtonian
behavior in the form of overall decrease of 
the attenuation coefficient and appearance of small oscillations
on the curve indicating that the wave has entered the 
non-Newtonian regime.
For $\alpha=1$, the attenuation coefficient
settles at a
somewhat lower value, and this happens
already for smaller frequencies than in the
case of Newtonian fluid. Also, much more
pronounced oscillatory structure of the solution
can be noticed.

\section{Numerical Results for Propagation of Dilatational Waves}

We calculated normalized phase velocities
of the plane dilatational waves, $v_I/V_c$ and 
$v_{II}/V_c$, and the attenuation coefficients $L_c/x_{I}$
and $L_c/x_{II}$   using 
our more general expression for $F(\kappa)$ 
(Maxwell fluid filling the pores)
given by Eq. (15).

In Fig. 3 we plot phase velocity 
$v_I/V_c$ as a function of frequency for the 
case of $\alpha \to \infty$, in order to recover the
Newtonian limit obtained by Biot.
Note that this case  reproduces
curve 1 in Fig. 11 from  (Biot, 1956b). 

Fig. 4 shows phase velocity 
$v_I/V_c$ as a function of frequency for the case of
$\alpha=1$, corresponding to the deeply
non-Newtonian regime. We notice again appearance of an 
oscillatory structure of the solution.
Also, phase velocity  $v_I/V_c$
settles at a
somewhat higher value 
than in the Newtonian case, and this happens
already for smaller frequencies.

In Fig. 5 we plot phase velocity 
$v_{II}/V_c$ as a function of frequency for the 
cases of $\alpha \to \infty$ and $\alpha =1$. 
Note that the case of  $\alpha \to \infty$ (thick solid line) 
perfectly reproduces the Newtonian limit obtained by Biot
curve 1 in Fig. 12 from  (Biot, 1956b). 
For $\alpha=1$, again we notice an
oscillatory structure of the solution.
Besides, we observe that the phase velocity  $v_{II}/V_c$
settles again at a
somewhat higher value than in the Newtonian case.

In Fig. 6 we plot the attenuation coefficient  
$L_c/x_I$  as a function of frequency for the three
cases: the thick  curve corresponds to $\alpha \to \infty$
(Newtonian limit), the dashed  curve
corresponds to a slightly sub-critical value of $\alpha=10$ 
and thin solid curve
corresponds to the case of $\alpha=1$.
The $\alpha \to \infty$ case  reproduces
curve 1 in Fig. 13 from  (Biot, 1956b). For $\alpha=10$,
we notice deviation from the classic Newtonian
behavior in the form of the overall increase of the attenuation
coefficient
 and appearance of small oscillations
on the curve, which indicates that the wave has entered the 
non-Newtonian regime. The large spike at 
low frequencies is also  due to
non-Newtonian effects.
For the case of  $\alpha=1$ the attenuation coefficient settles at
somewhat lower values, and this happens
already for smaller frequencies than in the
case of a Newtonian fluid. Also, a much more
pronounced oscillatory structure of the solution
can be noticed.

In Fig. 7 we plot the attenuation coefficient  
$L_c/x_{II}$  as a function of frequency for the three
cases: the thick curve corresponds to $\alpha \to \infty$
(Newtonian limit), the dashed  curve
corresponds to  $\alpha=10$ 
and thin solid curve
corresponds to the case of $\alpha=1$.
Note that the $\alpha \to \infty$ case perfectly matches
curve 1 in Fig. 14 from  (Biot, 1956b). For $\alpha=10$,
we  notice a deviation from the classic Newtonian
behavior in the form of the overall decrease in 
the attenuation coefficient and appearance of small oscillations
on the curve. The jump at $f/f_c=1$ (dashed curve) should be attributed to
the non-Newtonian effects.
For the case of 
$\alpha=1$ the attenuation coefficient settles again at
somewhat lower value and this happens
already for smaller frequencies than in the
case of the Newtonian fluid. Also, much more
pronounced oscillatory structure of the solution
can be noticed again.

\section{discussion}

In this paper, we have studied the
non-Newtonian effects in the propagation of elastic waves in porous
media by calculating phase velocities and attenuation
coefficients of the rotational and dilatational waves
as a function of frequency.
Originally, Biot (Biot, 1956a,b) performed similar analysis for
a Newtonian fluid-saturated porous medium.
Using our recent results [Paper I], and motivated
by a current need in models of propagation of elastic
waves in porous media with {\it realistic fluid rheologies},
we have generalized the work of Biot to the case of 
a non-Newtonian (Maxwell) fluid-saturated porous medium.

In summary, we found that replacement of an ordinary Newtonian fluid
by a Maxwell fluid in the fluid-saturated
porous medium results in
\begin{itemize}
\item an overall increase of the phase velocities of both the rotational  
and dilatational waves. With the increase of frequency these
quantities tend to a fixed, higher, 
as compared to the Newtonian limiting case, level
which is not changing with the decrease of the Deborah number
$\alpha$. 
\item the overall decrease of the attenuation coefficients of both the 
rotational and dilatational waves. With the increase of frequency these
quantities tend to a progressively lower,
as compared to the Newtonian limiting case, levels
as $\alpha$ decreases. 
\item Appearance of oscillations in all 
physical quantities 
in the deeply non-Newtonian regime when $\alpha \ll \alpha_c=11.64$.
\end{itemize}

The investigation of properties of  elastic
waves is  important for a number of
applications. The knowledge of phase velocities and
attenuation coefficients of  elastic waves in a 
realistic [such as saturated with Maxwell fluid] 
porous medium  
is necessary, for example,  to guide the oil-field exploration 
applications, acoustic stimulation of oil
producing fields (in order to increase the amount of recovered
residual oil), and the acoustic 
clean up of contaminated aquifers (Beresnev \& Johnson, 1994), (Drake \& Beresnev, 1999).

The idea of the paper was to use the function, $F(\kappa)$, 
that measures the deviation from Poisseuille flow friction,
extended to Maxwell fluids, and to substitute it into Biot's equations
of
poroelasticity without changing the latter.
However, Biot's equations have been derived under a number of 
assumptions. One
of these assumptions is that deviatoric (shear) components of the
macroscopic stress in the fluid are negligible Pride {\it et al.} (1992). 
 Pride {\it et al.} (1992) have shown that this assumption
is justified when $\eta \omega \ll G$, where $\eta$ is the dynamic 
viscosity of the fluid, $\omega$ is
frequency, and $G$ is frame shear modulus. 
Simple analysis shows that for typical Newtonian fluids such as water, 
this condition is only violated at frequencies $\omega > 10^9$ 1/s, 
or $f= \omega/(2 \pi) > 10^8$ Hz. 
Thus, for all frequencies below 1 MHz Biot's assumption is justified.
However, when we introduce the Maxwell fluid, the situation changes in
that
we introduce the real (in addition to imaginary) shear stresses. 
In summary, for any rheology (including Maxwellian) 
Biot's theory is only valid if macroscopic shear
stresses are negligible. In order to prove that, we
note from the rheological equation for a Maxwell fluid
$$
t_m {{\partial \tilde \tau}\over{\partial t}}= -\eta \nabla \vec  v  - 
\tilde \tau, 
$$
where $\tilde \tau$ represents the viscous stress tensor, that
in the frequency domain we can effectively obtain
$$
\tilde \tau = -\eta \nabla \vec  v / (1 +i t_m \omega).
$$
This means that we can roughly replace $\eta$ in all our estimates
with $\eta' / (1 +i t_m \omega)$.
There are two limiting cases. When $\omega \ll 1/t_m$, then
the fluid is effectively Newtonian and Biot's theory is valid.
When $\omega \gg 1/t_m$, i.e. when the fluid is essentially
non-Newtonian,
we effectively have $\eta'=\eta/(i t_m \omega)$, which in this
case is smaller than $\eta$ in absolute value. Thus, when
substituted into the shear stress, $S$, it produces $S= i \eta' \omega=
\eta/t_m$, which is smaller than $\eta \omega$. Therefore, we
conclude that inequality $\eta' \omega \ll G$ still holds for the
Maxwellian
fluid, i.e. Biot's equations are valid for Maxwell rheology.

This study, similarly to the results of Paper I, 
has clearly shown the transition from dissipative to
non-Newtonian regime in which sharp oscillations 
of all physical quantities are found.
We would like to comment on these unexpected
strong oscillations that were demonstrated
by our numerical analysis.
The results are based on
equation (15), which has been derived for a circular cylindrical
geometry. This is the same geometry that was used in classical
works of Biot and others. For Newtonian fluids the use of such
an idealized geometry for porous materials was backed up by an
analysis that showed that the results are not very sensitive to
the particular geometry (see, e.g., Johnson {\it et al.} (1987)).
Of course, the magnitude of these
oscillations depends on fluid parameters and permeability, and
may not be as high for many fluids. However, even if parameters
are such that oscillations are large, it is unclear at this stage
whether this oscillatory behavior will hold for more realistic
geometry, i.e. when curved pore walls (tortuosity) are considered.
There is a possibility that with tortuosity effects included
the obtained  oscillations will be smeared. 
However, our goal was to constrain ourselves with
simple cylindrical geometry, and a
separate study is needed to analyze the tortuosity effects
on our results.

\begin{acknowledgements}

This work was supported by the Iowa State University Center for
Advanced Technology Development and ETREMA Products, Inc.

\end{acknowledgements}
\newpage
\centerline{\bf REFERENCES}
\vskip 1cm

 Beresnev, I.A., Johnson, P.A.: 1994,
Elastic-wave stimulation of oil production:
A review of methods and results,
{\it Geophys.}, {\bf 59}, 1000-1017

 Biot, M.A.: 1956a,
Theory of propagation of elastic waves in a fluid-saturated
porous solid. I. low-frequency range, {\it J. Acoust. Soc. Am.}, 
{\bf 28}, 168-178

 Biot, M.A.: 1956b,
Theory of propagation of elastic waves in a fluid-saturated
porous solid. II. higher-frequency range, {\it J. Acoust. Soc. Am.}, 
{\bf 28}, 179-191

 Carcione, J.M., Quiroga-Goode, G.: 1996, 
Full frequency-range transient solution for compressional
waves in a fluid-saturated viscoacoustic porous medium,
{\it Geophysical prospecting}, {\bf 44},
99-129

 del Rio, J.A., de Haro, M.L., Whitaker, S.: 1998, 
Enhancement in the dynamic response of a viscoelastic 
fluid flowing in a tube,
{\it Phys. Rev. E}, {\bf E58}, 6323-6327

 Drake, T., Beresnev, I.: 1999,
Acoustic tool enhances oil production, 
{\it  The American Oil \& Gas Reporter}, September issue,
p.101-104

 Johnson, D.L., Koplik, J., Dashen, R.: 1987, 
Theory of dynamic permeability and tortousity in a fluid-saturated
porous media,
{\it J. Fluid. Mech.}, {\bf 176}, 379-402

 Pride, S.R., Gangi, A.F., Morgan, F.D.: 1992, 
Deriving the equations of motion for porous isotropic
media, {\it J. Acoust. Soc. Am.},
{\bf 92}, 3278-3290

 Tsiklauri, D., Beresnev, I.: 2001,
Enhancement in the dynamic response of a viscoelastic 
fluid flowing through a longitudinally vibrating tube,
{\it Phys. Rev. E}, {\bf 63}, 
046304-1-4 (2001), Paper I

\newpage
\centerline{\bf Figure captions}

Fig. 1 Behavior of dimensionless, normalized phase velocity of
the rotational wave, $v_r/V_r$, 
as a function of frequency. The thick solid curve
corresponds to the Newtonian limit when $\alpha \to \infty$, while
dashed and thin solid curves represent the non-Newtonian cases 
 $\alpha = 10$ and $\alpha= 1$ respectively.

Fig. 2 Behavior of dimensionless, normalized attenuation coefficient of
the rotational wave, $L_r/x_a$, 
as a function of frequency.  The thick solid curve
corresponds to the Newtonian limit when $\alpha \to \infty$, while
dashed and thin solid curves represent the non-Newtonian cases 
 $\alpha = 10$ and $\alpha= 1$ respectively.

Fig. 3 Behavior of dimensionless, normalized phase velocity of
the dilatational wave, $v_I/V_c$,
as a function of frequency. 
Here the Newtonian limiting case,  when $\alpha \to \infty$, 
is plotted.

Fig. 4 Same as in Fig. 3 but for $\alpha= 1$.

Fig. 5 Behavior of dimensionless, normalized phase velocity of
the dilatational wave, $v_{II}/V_c$, 
as a function of frequency. 
Here the Newtonian limiting case,  when $\alpha \to \infty$, 
is plotted with thick curve, while the thin one corresponds to the
non-Newtonian case of $\alpha= 1$.

Fig. 6 Behavior of dimensionless, normalized attenuation coefficient of
the dilatational wave, $L_c/x_I$, 
as a function of frequency. The thick solid curve
corresponds to the Newtonian limit when $\alpha \to \infty$, while
dashed and thin solid curves represent the non-Newtonian cases 
 $\alpha = 10$ and $\alpha= 1$ respectively.

Fig. 7 Behavior of dimensionless, normalized attenuation coefficient of
the dilatational wave, $L_c/x_{II}$, 
as a function of frequency. The thick solid curve
corresponds to the Newtonian limit when $\alpha \to \infty$, while
dashed and thin solid curves represent the non-Newtonian cases 
 $\alpha = 10$ and $\alpha= 1$ respectively.

\end{article}
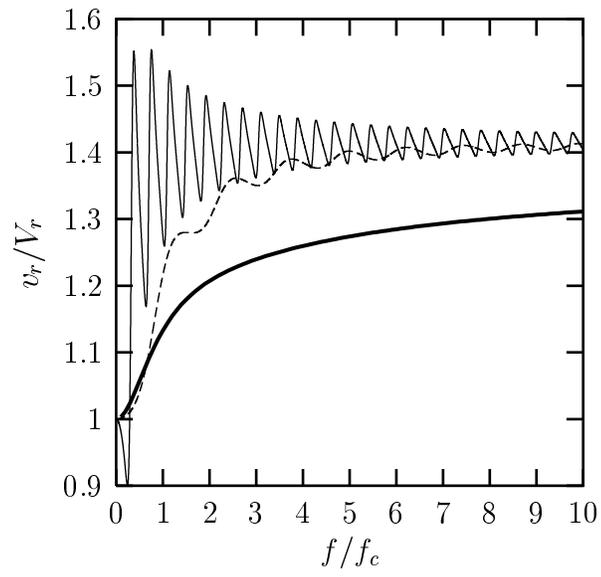
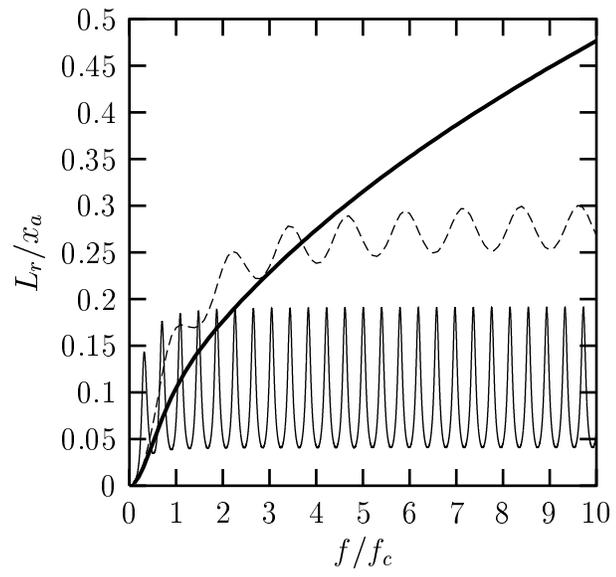
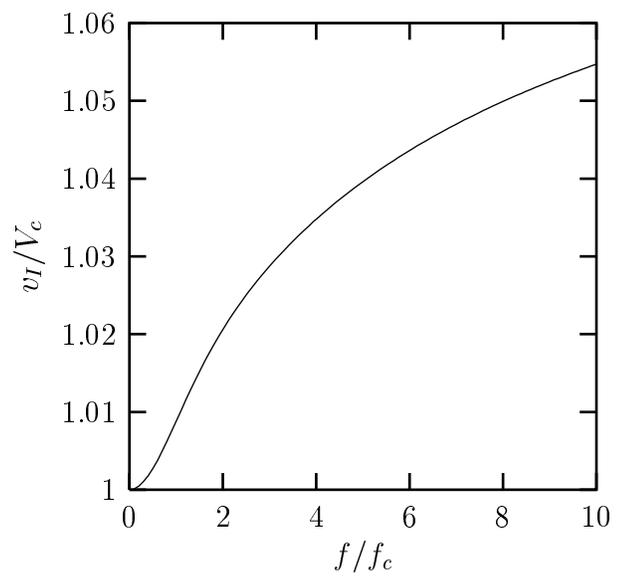
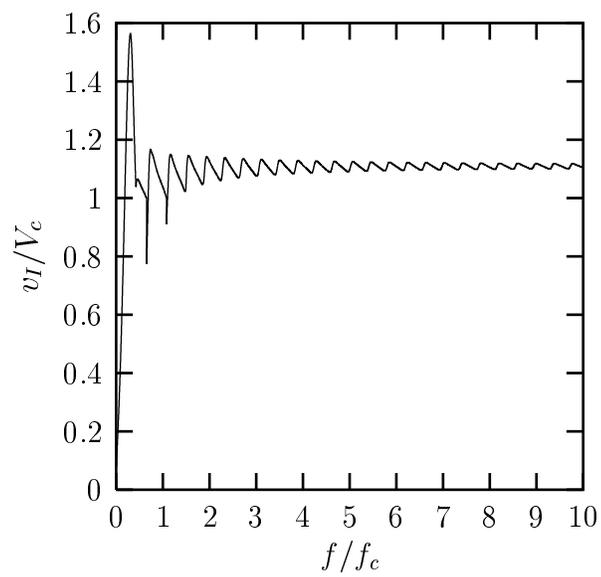
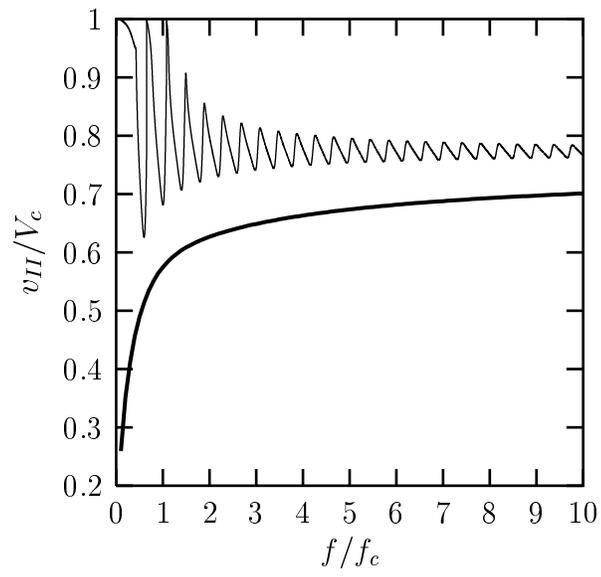
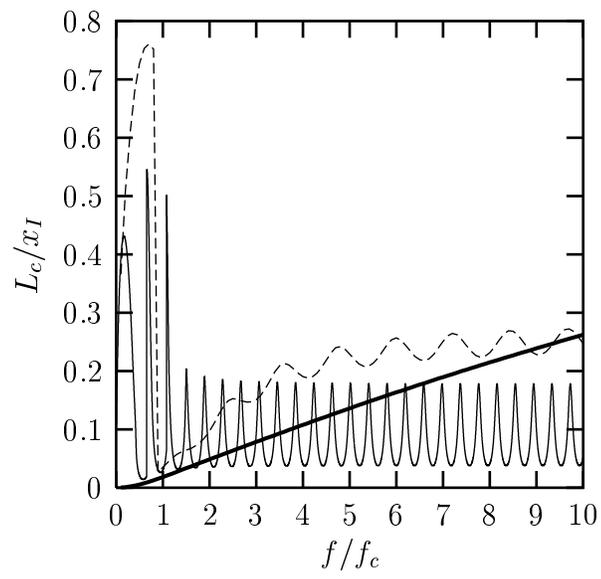
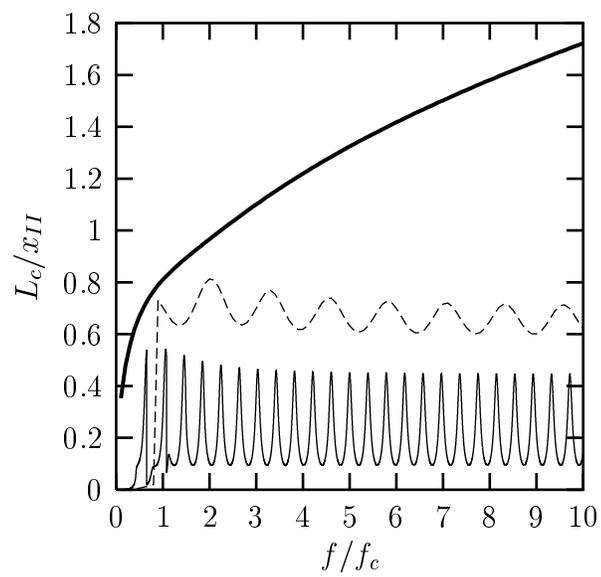
\end{document}